\documentclass[runningheads,pdflatex,sn-mathphys-num]{llncs}

\usepackage{graphicx}
\usepackage{multirow}
\usepackage{amsmath,amssymb,amsfonts}
\usepackage{mathrsfs}
\usepackage[title]{appendix}
\usepackage{xcolor}
\usepackage{textcomp}
\usepackage{manyfoot}
\usepackage{booktabs}
\usepackage{algorithm}
\usepackage{caption}
\usepackage{amsmath,amssymb}
\usepackage{algorithmicx}
\usepackage{algpseudocode}
\usepackage{listings}
\usepackage{makecell}
\usepackage{tabularx}
\usepackage{array}
\usepackage{soul}
\usepackage{enumitem}
\setlist[itemize]{
    itemsep=0pt,
    topsep=1pt,
    parsep=0pt,
    partopsep=0pt
}
\usepackage[backend=bibtex,style=numeric,sorting=none]{biblatex}
\begin{document}
\title{An Enhanced Adaptive Graph Neural Network (AGNN) for Hardware Trojan Identification}
\author{Banalata Bhunia\inst{1} Malay Kule\inst{2} Hafizur Rahaman\inst{3}}

\institute{
Indian Institute of Engineering Science and Technology, Shibpur,\\
Howrah, West Bengal, India \inst{1,2,3}\\
School of VLSI Technology\inst{1},\\
Department of Computer Science and Technology\inst{2},\\
Department of Information Technology\inst{3},\\
\email{bhuniabanalata@gmail.com\inst{1}}\\
\email{malay.kule@gmail.com\inst{2}}\\
\email{hafizur@vlsi.iiests.ac.in\inst{3}}
}
\titlerunning{AGNN for Hardware Trojan Identification}
%
%
%
%
\maketitle              

\begin{abstract}
Integrated circuits (ICs) face increasingly critical security challenges as semiconductor design and manufacturing become more globally distributed. Malicious modifications, particularly Hardware Trojans (HTs), may be introduced during fabrication, potentially compromising system confidentiality, integrity, and reliability. Conventional detection techniques often struggle to identify stealthy and structurally complex HTs due to limited scalability and inadequate feature representation. To address these challenges, this paper proposes an Adaptive Graph Neural Network (AGNN)-based framework for Hardware Trojan detection and implementation analysis. Experimental evaluation demonstrates an average accuracy of 99.95\%, precision of 99.06\%, recall of 99.59\%, and F-measure of 99.29\%. Furthermore, comparisons with state-of-the-art techniques indicate that the proposed framework achieves competitive detection performance across key evaluation metrics. These results highlight the potential of the proposed framework as a scalable, learning-driven solution for strengthening hardware security and facilitating the design of trustworthy and reliable integrated circuits.


\end{abstract}
\vspace{-10pt}
\section{Introduction}
\begin{figure}[!t]
\centering
\includegraphics[width=0.80\textwidth]{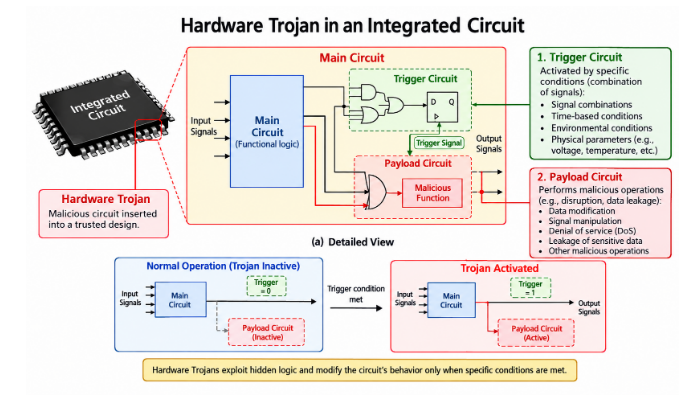}
 \vspace{-5pt}
\caption{Trojan inserted Integrated Circuit (IC)}
\label{fig:Trojan_in_ic}
\end{figure}

Modern electronic systems rely on third-party intellectual property cores, and external manufacturers provide semiconductor fabrication. The potential of machine learning (ML) and deep learning (DL) approaches has been demonstrated by recent developments in hardware security applications \cite{b16,b17,b19,b26,b27,b28}. Now a days, Hardware Trojan(HT) identification at gate-level using Graph Neural Networks (GNNs) are popular \cite{b13}\cite{b14}. A unique coding technique that combines logic gate type and port connection information is developed in Golden-free Trojan-gate identification technology \cite{b10}\cite{b11}. Coding method that integrates logic gate type and port connection data is developed for circuit graph modeling \cite{b9}\cite{b24}. An automated GNN detection architecture based on Bi-directional Graph Convolutional Network (Bi-GCN)\cite{b12}is developed to aggregate both the circuit signal propagation structure features from the circuit graph\cite{b15}\cite{b25}.   Gate-level  HT identification technique over 10,000 nets utilizing random forests and neural networks machine learning techniques \cite{b18}\cite{b21}. The features that are extracted for each net in a given netlist are learned by the neural networks and random forests \cite{b20}. For fast and accurate data gathering, Artificial Intelligence and Machine Learning accelerators are used to proposed a segmentation approach tailored for processing PSCA data that demonstrate the actual power measurement \cite{b23}\cite{b21}.Side channel analysis is also used to characterizes Trojan’s behavior that includes trigger type, payload type and bottom-up approach that combined real power trace and ML technique \cite{b22}\cite{b25}.
The fundamental idea of a Hardware Trojan (HT) and its workings are shown in Fig.~\ref{fig:Trojan_in_ic}. 
For gate-level netlists, a hybrid Hardware Trojan detection method based on the XGBoost algorithm is suggested \cite{b4}\cite{b3}\cite{b2}. The average positive rate increases to 84.6\% and the average true negative rate to 95.1\%, according to the experimental data. Adaptive Synthetic Sampling Technique (ADASYN) is used to normalize the extraction data, and supervised algorithms \cite{b5}. Sequential and combinational logic of Trojan’s structural features are extracted to construct an HT feature database \cite{b6}. Research on Trust-Hub benchmarks suggests that the suggested approach might detect any stealth Trojan with minimal runtime.\\
LMDet is a novel method for identifying HTs that uses natural language processing technology to distinguish between the `unnaturalness' of HTs and the `naturalness' of regular circuits \cite{b7}. LMDet has a reasonable space overhead and operates fast on large commercial circuits. Hardware Trojan (HT) identification in machine learning using SVM techniques is also successful \cite{b15}.  GramsDet is a unique technique that uses recurrent neural networks to record suspicious circuit connection topologies in order to identify HT \cite{b8}. An order-sensitive co-occurrence matrix is used for `gate embedding' and n-gram circuit for modeling the target circuit.     \\
In the proposed framework, benchmark Trust-HUB circuits are employed for detecting Hardware Trojans (HT) via an Adaptive Graph Neural Network (AGNN). Gate-level netlists are generated and then transformed into data flow graphs ( DFG), where logic gates and signal connections are represented as nodes and edges. Then the graph dataset is split into test, train and validation. AGNN is applied in graph dataset where AGNN introduces an adaptive aggregation mechanism that determines the relative relevance of nearby nodes. This allows the model to suppress less important information and concentrate on crucial structural relationships connected to Trojan-infected areas. Experimental results show that the suggested AGNN provides better detection performance, greater true positive and true negative rates and improved robustness in identifying sparse and stealthy Hardware Trojans when compared to current state-of-the-art GNN based approaches. \\
This paper is structured such that Section 2 serves as the introduction to the proposed work. The proposed hardware Trojan detection model, described in Sections 3 and experimental results is presented in section 4. Section 5 describes the advantages and challenges of the Proposed Adaptive Graph Neural Network (AGNN). Section 6 concludes the paper, summarizing the work completed so far and presenting future plans. 

\section{Proposed Method of Hardware Trojan(HT) Identification}
\subsection{Netlist Parsing}
Netlist parsing involves retrieving the structural details from analyzing the Verilog code of a digital circuit.  In this phase, the parser finds every logic element in the design, including flip-flops, logic gates, and multiplexers. Each gate is linked with the input and output wire.  These connections are employed to determine each node’s fan-in and fan-out properties, with fan-in representing the count of incoming signals to a gate and fan-out indicating the number of outputs it drives.  

\vspace{-5pt}
\subsection{Graph Construction}
\begin{figure}[!t]
\centering
\includegraphics[width=0.75\textwidth]{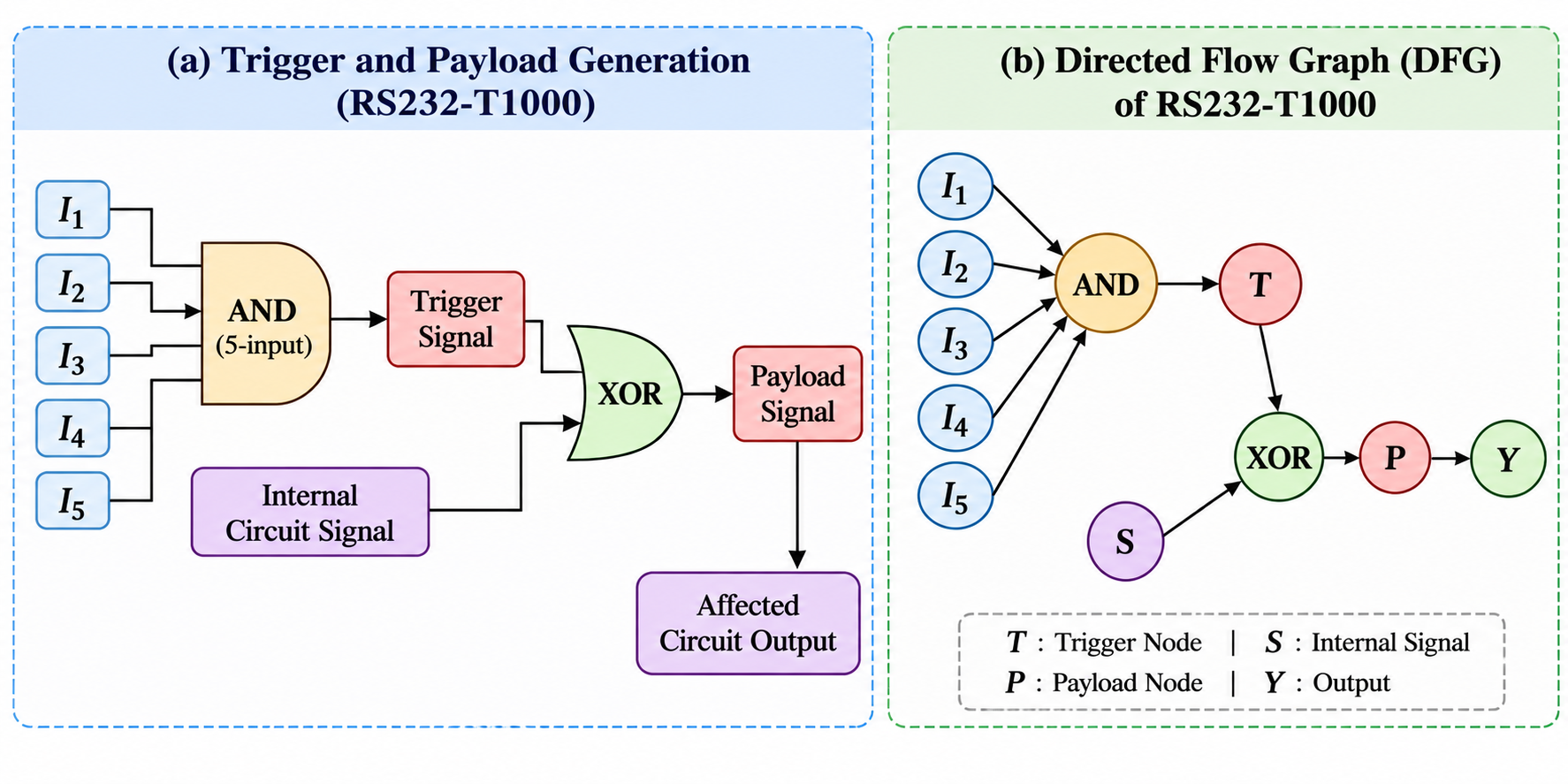}
 \vspace{-5pt}
\caption{Data Flow Graph (DFG) representation of RS232-T1000 Trojan Circuit}
\label{fig:DFG_representation}
\end{figure}

After the netlist parsing stage, the hardware circuit is converted into a directed data flow graph, which defined as $ DFG = ( V, E) )$ with  $V$  representing the nodes for logic gates, flip-flops, inputs, and outputs, and $ E $indicating the directed edges. 
Fig.~\ref{fig:DFG_representation} represents the circuit diagram and Data Flow Graph (DFG) of RS232-T1000 Trojan circuit. The circuit with the inserted Trojan includes extra trigger and payload functionality. A rare trigger condition is generated via a 5-input AND gate, which fires only when all primary inputs are set high. The trigger signal is then merged with an internal signal via an XOR gate to generate the payload, which alters the output of the circuit.

\vspace{-5pt}
\subsection{Problem Model}
This section offers a thorough description of the Adaptive Graph neural framework for HT detection and Trojan-gate categorization. The proposed methodology is predicated on the existence of a feedforward classification function $f_(i)$ , where the circuit is $i$,$j$identifies the gate/node within that circuit and mapping a hardware design p, to a binary state: infected or uninfected by a malicious Trojan circuit.
\begin{equation}
    f_(i, j)[p] = 
    \begin{cases} 
        (0,1) & \text{if Trojan exists }   \\
        (1,0) & \text{otherwise}
    \end{cases}
\end{equation}
To approximate the function, adaptive Graph Neural Network based hardware Trojan detection is introduced.

\vspace{-5pt}
\subsection{Feature Extraction of DFG}
Many features are extracted for implementation of HT, like structural, topological, Logical and Statistical. Gate type and Logic depth are two structural features. Fan-in , Fan-out and node degree, these three are topological features. Switching activity is statistical feature of the DFG. The frequency of logic transitions at a gate output while a circuit is operating is represented by switching activity. During logic simulation, it is approximated by watching each gate's output signal under applied input test patterns. The ratio of the number of observed output transitions to the total number of simulation cycles is used to determine a gate's activity factor. This feature helps differentiate between typical logic activity and anomalous switching behavior linked to Hardware Trojan circuitry by providing information about the dynamic behavior of circuit nodes. These are explained in Table ~\ref{tab:dfg_features}.
The switching activity of gate $P_i$ can be expressed as:
\vspace{-3pt}
 \begin{equation}
\small
Switching Activity(p_i) =
\frac{N_{\mathrm{transitions}}(p_i)}
{N_{\mathrm{total\ simulation\ cycles}}}
\label{eq:switching_activity}
\end{equation}
\vspace{-8pt}

where:
\begin{itemize}[nosep, leftmargin=15pt]
    \item $N_{\mathrm{transitions}}(p_i)$ is the number of output transitions ($0\rightarrow 1$ or $1\rightarrow 0$) observed at gate $p_i$.
    \item $N_{\mathrm{total\ simulation\ cycles}}$ is the total number of simulation cycles.
\end{itemize}
\vspace{-3pt}

\begin{table*}[t]
\centering
\caption{Features of Hardware Trojan Detection}
\label{tab:dfg_features}

\renewcommand{\arraystretch}{1.0}
\setlength{\tabcolsep}{4pt}

\begin{tabularx}{\textwidth}{%
>{\raggedright\arraybackslash}p{1.4cm}
>{\raggedright\arraybackslash}p{1.4cm}
>{\raggedright\arraybackslash}X
>{\raggedright\arraybackslash}X
}

\hline
\textbf{Feature of DFG} &
\textbf{Types of Features} &
\textbf{Description of Features} &
\textbf{Relevance of HT} \\
\hline

Gate type &
Structural &
Logic gate type &
Trojan circuitry address unnecessary gates \\

Fan-in &
Topological &
Incoming connection count &
Complication of local connectivity \\

Fan-out &
Topological &
Outgoing connection count &
Locating anomalous signal \\

Node degree &
Topological &
Connected neighbors count &
Unusual connectivity caused by Trojan logic \\

Logic depth &
Structural &
Distance of input/output nodes &
Describes the propagation of signals \\

Switching activity &
Statistical &
Transition frequency of the signal &
A Trojan node has different activity \\

\hline
\end{tabularx}
\end{table*}
\begin{algorithm}[t]
\captionsetup{labelformat=empty}
\caption{Algorithm: Hardware Trojan Detection Using an Adaptive Graph Neural Network (AGNN)}
\label{alg:agnn_detection}
\small

\begin{algorithmic}[1]

\Require Trust-HUB benchmark circuits
\Ensure Trojan-Free (0) or Trojan-Infected (1)

\State Load Trust-HUB circuits and insert Trojans using Trigger and Payload Logic
\State Construct dataset $D=\{C_1,C_2,\ldots,C_n\}$

\ForAll{$C_i\in D$}
    \State Convert $C_i$ into DFG $G_i=(V_i,E_i)$ and extract features $F_i$
    \State Assign label $Y_i=1$ if Trojan logic exists; otherwise $Y_i=0$
\EndFor

\State Split $D$ into $D_{\mathrm{train}}$ (70\%), $D_{\mathrm{val}}$ (15\%), and $D_{\mathrm{test}}$ (15\%)
\State Initialize AGNN and Adam optimizer

\For{$epoch=1$ to $MaxEpoch$}
    \ForAll{$G_i\in D_{\mathrm{train}}$}
        \State Apply adaptive graph attention and neighborhood message passing
        \State Generate node embeddings and obtain graph representation $H_i$ through graph pooling
        \State Compute class probabilities:
        \Statex \hspace{\algorithmicindent}$P_i=\operatorname{Softmax}(W_cH_i+b_c)$
        \State Calculate cross-entropy loss:
        \Statex \hspace{\algorithmicindent}$Z=-\sum_i\sum_{c=0}^{1}y_{ic}\log(P_{ic})$
        \State Update AGNN parameters using Adam optimizer
    \EndFor
    \State Validate model using $D_{\mathrm{val}}$
\EndFor

\State Evaluate trained AGNN on $D_{\mathrm{test}}$

\ForAll{$G_i\in D_{\mathrm{test}}$}
    \State Predict Trojan-Infected (1) if $P_i(1)>P_i(0)$; otherwise predict Trojan-Free (0)
\EndFor

\end{algorithmic}
\end{algorithm}

\vspace{-5pt}
\subsection{Architecture of the Proposed Model}
Benchmark circuits are obtained from the Trust-HUB repository in Verilog format, and Hardware Trojans (HTs) comprising trigger and payload logic are inserted to generate malicious samples. The resulting netlists are transformed into Data Flow Graphs (DFGs), where logic gates are represented as nodes and signal connections as directed edges. 
Features derived from the DFGs, including structural and topological ones, are utilized to build the graph dataset, it is further split into subsets for testing (15\%), validation (15\%), and training (70\%). A learning rate of 0.001, a batch size of 32, up to 200 epochs, a hidden dimension of 128, three AGNN layers, four attention heads, a dropout rate of 0.5, a weight decay of $5\times10^{-4}$, and a random seed set to 40 are used to train the AGNN model. Early stopping uses validation loss and waits up to 20 epochs before halting.
The "adaptive" characteristic of the proposed AGNN refers to dynamically learned neighborhood-aggregation weights during message passing.Each neighboring gate is weighted according to its learned relevance to the target gate, enabling the model to emphasize structurally significant relationships associated with potential Trojan regions. The proposed AGNN dynamically adjusts weights for neighborhood aggregation, prioritizing structurally significant neighboring gates to better detect potential Hardware Trojan areas. 



The Adaptive Graph Neural Network ( AGNN) is trained using the cross-entropy loss function, which measures the discrepancy between actual labels and predicted class probabilities.   
The loss function$(Z)$ is denoted as follows :
\begin{equation}
Z = -\sum_{k=1}^{M} \left[ x_k \log(q_k) + (1-x_k)\log(1-q_k) \right]
\end{equation}

\begin{figure}[!t]
\centering
\includegraphics[width=1.0\columnwidth]{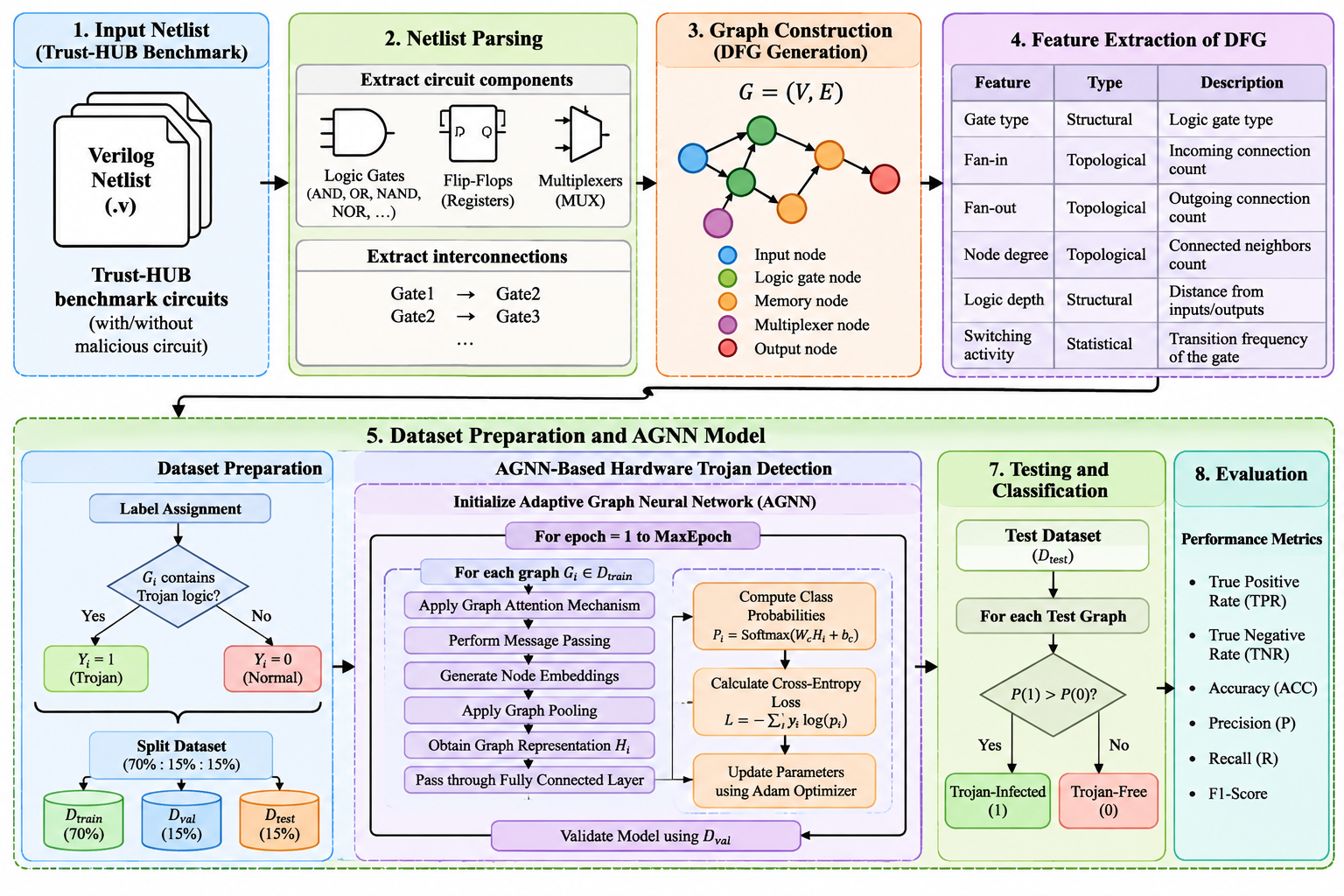}
\caption{Work Flow of Proposed Model }
\label{fig:architecture}
\end{figure}

True class label is denoted as $x_k$ and $q_k$ denotes the predicted probability for the $k$-th sample. 
{When $q_k \rightarrow 0$, we have}
{
\vspace{-3pt}
\begin{align}
(1-x_k)\log(1-q_k)
&\rightarrow (1-x_k)\log(1) \\
&= 0 \qquad \text{since } \log(1)=0
\end{align}
}
where sample number is denoted as $M$ , $k$ denotes sample index, $(1-x_k)$ and $(1-q_k)$ represent the corresponding negative class and its predicted probability, respectively. The model is optimized using Adam with adaptive learning rates. After the classification stage, a detailed Trojan analysis is performed to investigate the characteristics of malicious circuitry within the infected designs. 
Fig.~\ref{fig:architecture} illustrates the architecture of proposed Hardware Trojan(HT) detection framework.

\section{Experimental Results and Discussion}\label{sec3}
\subsection{Dataset (Trust-HUB Benchmark circuit)}\label{subsec2}
Trust-HUB benchmark circuit is used for proposed model \cite{b1}. Multiple benchmark circuits such as S27-T100, S5378-T200 S9234-T300, S13207-T100, and RS232-T1000 etc were chosen from the Trust-HUB dataset for this study. These benchmarks include deliberately embedded Hardware Trojans(HT) featuring different trigger and payload mechanisms, such as attacks that cause information leakage or alter functionality.

\vspace{-5pt}
\subsection{Experimental Setup}
The proposed Adaptive Graph Neural Network ( AGNN) framework is implemented using contemporary software tools and high-performance computing resources. Python was chosen as the primary programming platform for executing the whole framework. 
The circuit netlists were converted into graph formats and analyzed with NetworkX.  At least 16 GB of RAM was needed to process large benchmark circuits and graph datasets efficiently. The experiments were conducted on a workstation featuring a GPU and CUDA support to speed up the AGNN model's inference and training. 

\vspace{-10pt}
\subsection{Performance Evaluation}
The machine learning-based suggested model requires values for False Negatives ( FN), True Negatives ( TN), False Positives ( FP), and True Positives ( TP) to assess its performance. True Negative is the number of normal (Trojan-free) circuits correctly classified as Trojan-free. False Positive denoted the number of normal (Trojan-free) circuits incorrectly classified as Trojan-infected. False Negative indicates count of Trojan samples wrongly classified as Trojan-free circuit. True Positive indicates the count of Trojan circuits accurately identified.  
Various machine learning matrices are employed to assess the proposed model, including Accuracy ( ACC), Precision ( P), F-Measure ( F1) and Recall. 

\begin{table*}[!t]
\centering
\caption{Performance metrics of the proposed Adaptive Graph Neural Network (AGNN) model}
\label{tab:performance}

\renewcommand{\arraystretch}{1.00}
\setlength{\tabcolsep}{3pt}
\small

\begin{tabular}{|l|c|c|l|c|c|c|c|c|c|}
\hline
\multirow{2}{*}{\textbf{Benchmark}} &
\multirow{2}{*}{\makecell{\textbf{Standard}\\\textbf{Circuit}}} &
\multirow{2}{*}{\makecell{\textbf{Malicious}\\\textbf{Circuit}}} &
\multirow{2}{*}{\textbf{Trojan's Influence}} &
\multicolumn{6}{c|}{\textbf{Proposed Method}}\\
\cline{5-10}
&&&&
\textbf{TN} & \textbf{FP} & \textbf{FN} &
\textbf{TP} & \textbf{TPR} & \textbf{TNR}\\
\hline

RS232-T1000 & 202 & 13 & Data contamination & 202 & 0 & 1 & 12 & 0.9230 & 1.0000\\
RS232-T1100 & 204 & 12 & Data contamination & 204 & 0 & 0 & 12 & 1.0000 & 1.0000\\
RS232-T1200 & 202 & 14 & Data contamination & 201 & 1 & 0 & 14 & 1.0000 & 0.9950\\
RS232-T1300 & 204 & 9 & Data contamination & 204 & 0 & 0 & 9 & 1.0000 & 1.0000\\
RS232-T1400 & 202 & 13 & Data contamination & 202 & 0 & 0 & 13 & 1.0000 & 1.0000\\
RS232-T1500 & 202 & 14 & Data contamination & 202 & 0 & 0 & 14 & 1.0000 & 1.0000\\
RS232-T1600 & 202 & 12 & Data contamination & 202 & 0 & 0 & 12 & 1.0000 & 1.0000\\
s35932-T100 & 5426 & 15 & Functional mod. & 5426 & 0 & 0 & 15 & 1.0000 & 1.0000\\
s35932-T200 & 5422 & 16 & Functional mod. & 5422 & 0 & 0 & 16 & 1.0000 & 1.0000\\
s35932-T300 & 5426 & 36 & Functional mod. & 5426 & 0 & 0 & 36 & 1.0000 & 1.0000\\
s38417-T100 & 5329 & 12 & Functional mod. & 5329 & 0 & 0 & 12 & 1.0000 & 1.0000\\
s38417-T200 & 5329 & 15 & Functional mod. & 5329 & 0 & 0 & 15 & 0.9375 & 1.0000\\
s38417-T300 & 5358 & 15 & Functional mod. & 5358 & 0 & 0 & 15 & 1.0000 & 1.0000\\
\hline

S27-T100 & 17 & 4 & Functional mod. & 17 & 0 & 0 & 4 & 1.0000 & 0.8823\\
S5378-T200 & 2914 & 104 & Functional mod. & 2914 & 0 & 1 & 103 & 0.9903 & 1.0000\\
S9234-T300 & 6983 & 78 & Functional mod. & 6983 & 0 & 0 & 78 & 1.0000 & 1.0000\\
S13207-T100 & 9577 & 247 & Functional mod. & 9577 & 0 & 0 & 247 & 1.0000 & 1.0000\\
S15850-T100 & 2155 & 27 & Functional mod. & 2155 & 0 & 0 & 27 & 1.0000 & 1.0000\\
S38584-T100 & 6473 & 9 & Functional mod. & 6472 & 1 & 0 & 9 & 0.7777 & 0.9998\\
S38584-T200 & 6473 & 83 & Functional mod. & 6472 & 0 & 0 & 83 & 1.0000 & 0.9998\\
S38584-T300 & 6473 & 731 & Functional mod. & 6473 & 0 & 0 & 731 & 1.0000 & 1.0000\\

\hline
\multicolumn{8}{|r|}{\textbf{Average}} &
\textbf{99.59\%} &
\textbf{99.98\%}\\
\hline
\end{tabular}
\end{table*}

\vspace{-10pt}
\subsection{Experimental Results}
AGNN-based Hardware Trojan(HT) detection framework achieves high detection accuracy for Trojans and strong performance in classifying normal gates across both small and large benchmark circuits. The model attains average TPR and TNR values both are exceeding 99\% , demonstrating its effectiveness in accurately detecting malicious Trojan in circuitry while reducing false alarms. The detail of the outcomes of the implemented model is shown in Table \ref{tab:performance}.

\begin{table*}[!t]
\centering
\caption{Comparison of the proposed model with existing Hardware Trojan (HT) detection methods}
\label{tab:comparison}

\renewcommand{\arraystretch}{0.75}
\setlength{\tabcolsep}{4pt}
\small

\begin{tabular}{|p{1.8cm}|p{1.5cm}|p{3.0cm}|c|c|c|}
\hline

\textbf{Paper (Year)} &
\textbf{Class. Item} &
\textbf{Technology-Method} &
\textbf{TPR} &
\textbf{TNR} &
\textbf{ACC} \\
\hline

(2019) \cite{b6} & Circuit & Threshold-Template library & NIL & NIL & 78\% \\
\hline
(2018) \cite{b7} & Path & ML-Language model & 96.5\% & 97.7\% & NIL \\
\hline
(2019) \cite{b8} & Path & ML-Recurrent neural network & 80.7\% & 95.5\% & NIL \\
\hline
(2020) \cite{b3} & Net & ML-Neural network & 84.6\% & 95.1\% & NIL \\
\hline
(2020) \cite{b4} & Net & ML-XGBoost hybrid detection & 90.9\% & 99.0\% & NIL \\
\hline
(2019) \cite{b5} & Net & ML-Multi-model & 82.46\% & 98.99\% & 98.26\% \\
\hline
GNN4Gate \cite{b2} & Gate & ML-Graph neural network & 87.14\% & 99.73\% & 99.61\% \\
\hline

\textbf{Proposed} &
\textbf{Gate} &
\textbf{ML-Adaptive GNN} &
\textbf{99.59\%} &
\textbf{99.98\%} &
\textbf{99.95\%} \\
\hline

\end{tabular}
\end{table*}

\begin{figure}[!t]
\centering
\includegraphics[width=1.0\columnwidth,height=6cm]{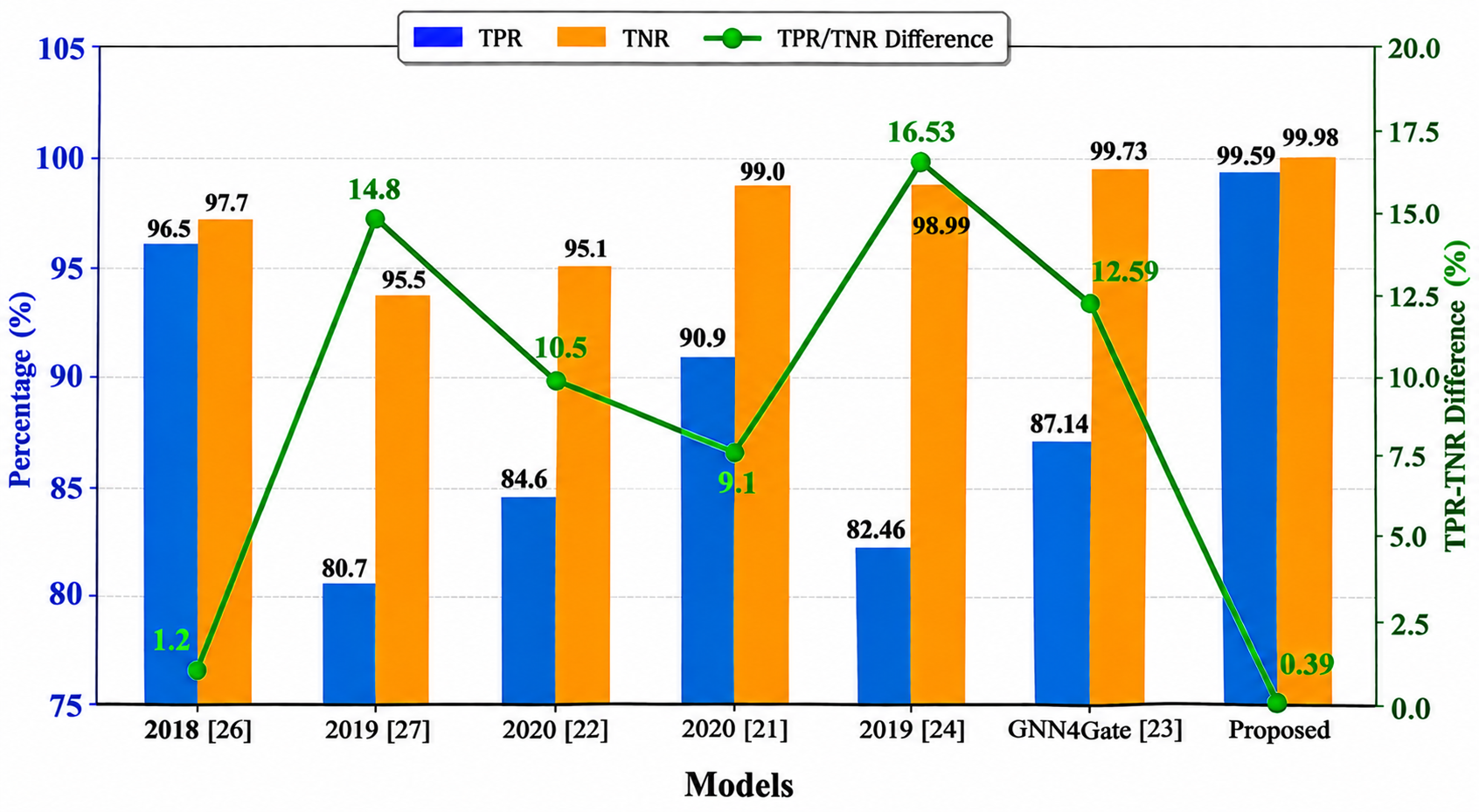}
\caption{Evaluation of the proposed AGNN model's performance corresponding to existing Hardware Trojan(HT) detection techniques}
\label{fig:Comparision}
\end{figure}
\subsection{Comparison of the AGNN based HT detection technique with the previous state-of-the-art techniques}

The comparison analysis displays in Table \ref{tab:comparison}.  Hardware Trojan(HT) detection methods like the Threshold-Based Template Library approach (2019)\cite{b6} classified Trojans at the circuit level but reached only 78\% accuracy whereas the Language Model-based approach ( 2018) \cite{b7} attained a TPR of 96.5\% and a TNR of 97.7\%.  The Recurrent Neural Network ( RNN) method ( 2019) \cite{b8} recorded a lower TPR of 80.7\% but still achieved a high TNR of 95.5\%. Later net-level detection methods such as Neural Networks (2020) \cite{b3}, XGBoost Hybrid Detection ( 2020) \cite{b4}, and Multi-Model Learning ( 2019) \cite{b5} reported true positive rates 84.6\%,90.9\% and 82.46\% , along with true negative rates near 99\%. While these methods decreased false alarms, their ability to detect Trojans stayed comparatively low. The recently proposed GNN4Gate \cite{b2} applied graph-based learning to detect Hardware Trojans at the gate level. By leveraging circuit topology data, it attained a TPR of 87.14\%, a TNR of 99.73\%, and an accuracy of 99.61\%, showcasing the efficacy of graph neural networks in this area. In comparison to GNN4Gate, the proposed AGNN boosts the Trojan detection rate from 99.61\% to 99.95\%, marking a substantial improvement in identifying malicious gates. The TPR rate is 99.59\% and TNR rate is 99.98\%, that is also higher than GNN4Gate,suggesting that false positives and false negative continue to be exceptionally rare. Experimental results demonstrates an average  precision of 99.06\%, recall of 99.59\%, and F-measure of 99.29\%. Furthermore, the high Precision and F-Measure values suggest that the AGNN based HT detection model effectively balances detection evaluation with classification accuracy.  Fig.~\ref{fig:Comparision} provides a comparative evaluation of the suggested method against previous HT detection techniques.  \\
The proposed adaptive graph neural network (AGNN) attains an Average True Positive Rate ( TPR) of 99.59\% and an Average True Negative Rate ( TNR) of 99.98\%, resulting in a TNR  and TPR gap of just 0.39\%.  A smaller difference between TNR and TPR suggests a more balanced classification performance for both Trojan-infected and Trojan-free circuit samples, reducing detection bias and improving the overall reliability of the model. 

\section*{Conclusion}

This work presents a Hardware Trojan (HT) detection framework based on an Adaptive Graph Neural Network (AGNN). 
The proposed method achieves high detection accuracy and a higher true positive rate while reducing false positives, demonstrating its effectiveness in distinguishing Trojan-infected circuits from Trojan-free designs. These findings establish the potential of the proposed AGNN-based framework as a scalable and efficient solution for hardware security analysis. Future work will focus on extending the framework to support real-time Hardware Trojan detection in complex, large-scale integrated circuits.

\subsubsection{Acknowledgements} This work has been partially supported by Ministry of Electronics and Information Technology, Government of India, under Chip to Statrtup Programme, vide Grant No. EE-9/2/2021-R\&D-E.

\printbibliography
\end{document}